# IRRADIATION OF LUMINESCENCE DOSIMETERS IN PULSED MIXED RADIATION FIELDS

*A. Cimmino\*,a, I. Ambrožová b, S. Motta a,c, R. Versaci a, V. Olšovcová a, R. Truneček a, A. Velyhan a, D. Chvátil d, V. Olšanský d, V. Stránský a, J. Šolc e*

a ELI Beamlines, Institute of Physics of CAS, Za Radnici 835, 25241 Dolni Brezany, Czech Republic
b Dep. of Radiation Dosimetry, Nucl. Phy. Inst. of CAS, Na Truhlarce 39/64, 18000 Prague, Czech Republic
c Paul Scherrer Institut, Forschungsstrasse 111, 5232 Villigen PSI, Switzerland
d Dep. of Accelerators, Nucl. Phy. Inst. of CAS, Husinec-Rez 130, 25068 Rez, Czech Republic
e Czech Metrology Institute, Okruzni 31, 638 00 Brno, Czech Republic

ABSTRACT

UHDpulse - Metrology for Advanced Radiotherapy using beams with Ultra-High Pulse Dose Rates is a European project aimed at developing novel dosimetry standards, as well as improving existing ones, for FLASH radiotherapy, very high energy electrons radiotherapy, and laser-driven medical accelerators. Within the scope of this project, Thermoluminescence (TL) and Optically Stimulated Luminescence (OSL) detectors are used to measure stray radiation fields. Experiments performed with conventional pulsed particle-beams allow to characterize the dosimeters in known and controllable radiation fields. In turn, this allows to develop models and predict their behavior in complex radiation fields, such as those at laser-driven and FLASH facilities. TL and OSL detectors were irradiated at the Microtron MT25 electron accelerator in Prague, Czech Republic. GAFChromic$^{TM}$ films and plastic nuclear track detectors were used to study the beam profile and the neutron background respectively. The responses of the different detector to the pulsed mixed radiation fields of the Microtron MT25 are compared among each other and presented in this paper.

## 1. Introduction

FLASH radiotherapy (FLASH-RT) [1] is a modern and promising cancer treatment that involves delivering the total prescribed dose almost instantaneously (~100 ms) at ultra-high dose rates ($10^6$ Gy/s) in ultra-short pulses (order of µs). This results in a significant reduction of adverse effects to the surrounding healthy tissue without compromising the treatment effectiveness [2, 3, 4, 5], i.e. the FLASH effect. The majority of the pre-clinical trials have been performed using electron beams [6, 7], but the FLASH effect was also observed in photons [8] and protons [9]. In the case of electrons, to allow for deep-seated tumor treatment, beams of 100 MeV or higher are needed. These energies are outside the reach of typical medical electron accelerators. In this context, the use of very high energy electrons (VHEE) beams [10] can address this limitation, but unfortunately bulky and expensive machinery is needed to achieve FLASH regime dose rates in conventional accelerators [11, 12]. On the other hand, laser-driven accelerators are seen as compact and cost-effective solutions for VHEE radiotherapy [13, 14, 15]. Still, before any future therapeutic application is possible, a solid and reproducible method for absorbed dose measurements at ultra-high pulse dose rates is required. Unfortunately, at laser-driven accelerators this is easier said than done. In fact, important metrological challenges are present due to the significantly high dose rates ($10^9$ Gy/s) and the ultra-short pulsed (up to ps-ns for protons and tens of fs for electrons) nature of the radiation fields involved [16]. In these conditions, the usual monitoring devices, such as ionization chambers or diodes, fail as their detection efficiency drops [17].

UHDpulse - Metrology for Advanced Radiotherapy using Particle beams with Ultra-High Pulse Dose Rates [16] is a joint European research project aimed at developing novel dosimetry standards, as well as improving existing ones, for FLASH-RT, VHEE radiotherapy, and laser-driven medical accelerators. The UHDpulse project aims at addressing the metrological issues associated with FLASH-RT described above by developing a metrological framework for traceable absorbed dose measurements in ultra-high pulsed dose rates particle beams.

\* Corresponding author
E-mail address: anna.cimmino@eli-beams.eu

Testing of personal and area dosimeters for operation in pulsed radiation fields is necessary to judge their suitability. This has multidisciplinary implications, as pulsed radiation fields are used in research, industry, security screening, medical treatments, and medical examinations. Furthermore, preliminary validation of dosimetry methodologies must be performed in reference or known radiation fields at conventional accelerators. Therefore, a dedicated data-taking campaign with OSL and TL detectors was performed at the Microtron MT25, a cyclic electron accelerator in Prague, Czech Republic. A relative comparison of the detectors was carried out and the results are presented in this paper.

## 2. Detector Systems and Methods for Dosimetry Outside Primary Beams

Within the UHDpulse project, focus is put on the development of traceable and validated methods for characterizing stray radiation fields. In fact, all therapeutic radiation beams studied in the project suffer from the generation of secondary radiation fields, which deliver unwanted dose to healthy tissue and organs outside the targeted area. Therefore, it is of the highest importance to be able to map secondary radiation fields. This is less straightforward than it seems. These fields have the same pulsed time structure of the primary beams. Additionally, they are mixed radiation fields, composed of different types of particles with different energies. Therefore their characterization is complex, since each field components needs to be identified.

While novel active detectors are being developed and successfully tested for this purpose [18, 19], passive solid-state dosimeters are being investigated as well. Optically stimulated luminescence and thermoluminescence detectors are ideal choices. These passive detectors have many advantages for these types of applications. They are robust and relatively inexpensive. They can easily be adapted to be placed in vacuum and comply with clean room environment. This is not usually required in radiotherapy, but it is an important aspect when laser optics are involved. Compared to active systems, there is no electronics that requires shielding from electromagnetic pulses present at laser-driven accelerators. Most importantly, they were already successfully tested in pulsed field with ultra-high dose rates [20, 21].

### 2.1. BeO Optically Stimulated Luminescence Detectors

Optically Stimulated Luminescence (OSL) [22] is the emission of light from a previously irradiated material when stimulated with photons of a specific energy. This results in a time dependent decay of luminescence where the intensity of the signal is an indirect measure of the absorbed radiation dose. Beryllium oxide optically stimulated luminescence (BeO-OSL) detectors are amongst the most used passive detectors exploiting luminescence phenomena. BeO-OSL detectors were chosen for this study based on their favorable dosimetric characteristics, which include a high sensitivity to photon ionizing radiation, a linear dose response over six orders of magnitude (µGy to Gy), and a low effective atomic number ($Z_{eff}$ = 7.2), which makes it a near tissue-equivalent material [23].

The BeO-OSLs used in this study were calibrated with a 137Cs at an accredited ISO/IEC 17025:2017 calibration laboratory [24]. Annealing was performed in air atmosphere at 700 ºC for 15 minutes in a laboratory chamber furnace by LAC s.r.o. Readout was done using a Lexsyg Smart reader from Freiberg Instruments GmbH. Each chip is stimulated for 120 s using blue light (wavelength 458 nm). In these conditions, > 99% of the signal is read at the current stimulation power of 85 mW/cm2. The signal is corrected for background and signal sensitivity to account for different production batches [25].

### 2.2. Thermoluminescence Detectors

Thermoluminescence (TL) [26] is a phenomenon of light emission caused by heating a pre-irradiated insulator or semiconductor. The intensity of the emitted light, also in this case, is proportional to the absorbed energy. A well-known TL phosphor, with excellent dosimetric properties, is lithium fluoride doped with magnesium, copper, and phosphorus (LiF:Mg,Cu,P) [27]. Moreover, building this phosphor using different Li isotopes creates distinct thermal neutron responses. In fact, Li-6 presents a much higher neutron capture cross section than Li-7. During the 1980s, a new technology of producing LiF:Mg,Cu,P dosimeters in the form of solid sintered pellets was developed [28]. These pellets with Li-6 and Li-7 are commercially known by their codes MCP-6 and MCP-7 [29, 30]. The main dosimetric properties of the MCP chips are available at https://www.radcard.pl. Annealing was performed at 240 ºC for 10 minutes followed by rapid cooling at room temperature. Preheating was done at 140 ºC for 15 s. The heating rate was 10 ºC/s and reading was at 240 ºC for 25 s.

Finally, dysprosium doped calcium sulfate ($CaSO_4$:Dy) phosphor is considered one of the most efficient phosphors for use in gamma and X-ray radiation dosimetry [31]. The CaSO4:Dy detectors used, were manufactured by Laboratories Protecta Ltd [32]. Preheating was done at 150 ºC for 22 s. The heating rate was 10 ºC/s and reading was at 280 ºC for 30 s.

For the MCP and CaSO4:Dy detectors, a Harshaw TLDTM Model 3500 was used for readout. The glow curve integral was evaluated. The measured signals were converted to absorbed dose via individual calibration for each detector (calibration source: 137Cs; calibration dose: 100 mGy).

## 2.3. CR-39s

Polyallyl diglycol carbonate ($C_{12}H_{18}O_7$) detectors, commercially referred to as CR-39s, were used as solid-state nuclear track detectors [33-34]. Specifically, Harzlas TD-1 (Nagase Landauer Ltd) with a thickness of 0.9 mm and density 1.3 g/cm$^3$ were used. After the irradiation, the detectors were etched in 5 N NaOH at 70 °C for 18 hours. The corresponding bulk etch layer was ~15 μm. The etched tracks were then analyzed using a high speed wide area imaging microscope HSP-1000. The absorbed dose and dose equivalent were calculated for heavy charged particles with linear energy transfer (LET) above 10 keV/μm (in water) by using the relationship between the parameters of etched tracks and LET of the particle. The energy threshold for neutron detection was about 100 keV. Neutrons were detected via secondary charged particles (recoil nuclei) formed as a result of interactions between the neutron and nuclei in the detector material or in the materials in close proximity to the detector.

## 2.4. GAFChromic$^{TM}$ HD-V Films

Few radiochromic films were available during the irradiation. These are self-developing films typically used in the testing and characterization of radiographic equipment. The dye in the film changes color when exposed to ionizing radiation. The color change provides a measurement of the absorbed dose, once the films are calibrated. The films used in this study were GAFChromic$^{TM}$ high-dose dosimetry (HD-V2) films [35]. They cover a wide range of charged-particle energies down to the keV level. Their operating dose range spans from 10 to 1000 Gy.

## 3. Experimental Setup

### 3.1. Microtron MT25

The Microtron MT25 [36] is a cyclic electron accelerator of the Nuclear Physics Institute of the Czech Academy of Sciences. Pulsed high-current monochromatic electron beams are produced and accelerated in a radio-frequency cavity. The electron beam has maximum energy of 25 MeV, pulse length of 3.5 μs, repetition rate of 423 Hz, and a mean maximum current of 30 μA. The electrons, emitted from a LaB$_6$ cathode, are accelerated in a Kapitza cavity resonator [35], in which an electric field with constant amplitude and frequency is applied. Subsequently, the energy of the electrons is gradually increased as it passes through the cavity resonator following circular trajectories of increasing radius in the vacuum chamber. Once the beam reaches the desired energy, it is extracted with an adjustable beam extractor and sent to the different workstations. For this study, the electron beam had a maximum energy of 23 MeV, an angular divergence of 12°, and maximum dose of 200 mGy/s. This is comfortably away from the FLASH-dose-rate regime. Therefore, the response of the detectors can be studied in a known and less demanding environment. This is needed and is preparatory for subsequent studies at FLASH electron facilities.

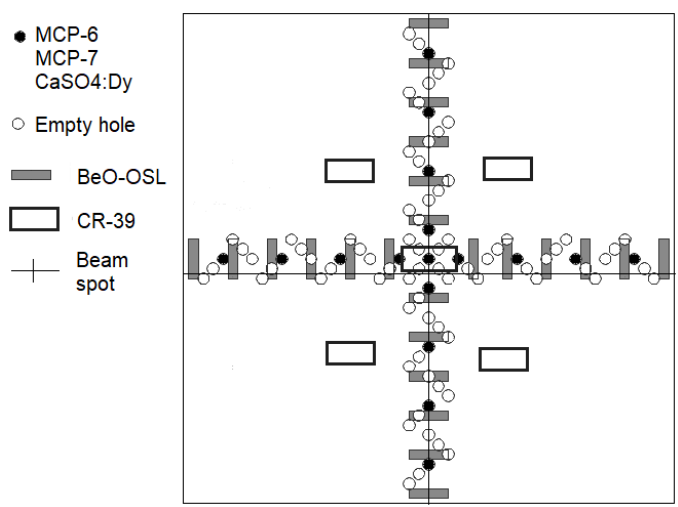

**Fig. 1 -** Schematic layout (not to scale) of the Plexiglas slab and radiation detectors. The MCP-6s, MCP-7s, and CaSO$_4$:Dys were stacked one on top of the other (filled-in black circles). 24 sets made of 4 BeO-OSL chips were wrapped in aluminium foil (filled-in grey boxes).

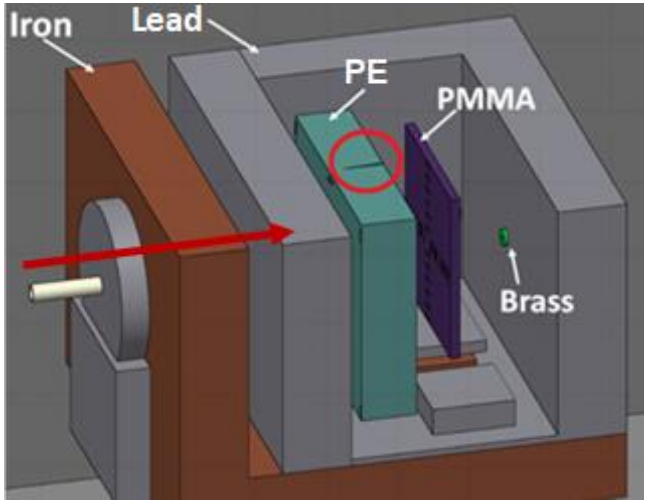

**Fig. 2** – 3D representation of the experimental setup with the PE moderator. The red circle indicates a crack in the moderator that has visible effects on the data, as shown in figure 4. The entrance collimator is positioned behind the iron wall and is not visible in this picture. The red arrow shows the beam direction.

### 3.2. Geometrical Setup

A lead and iron bunker (over 15 cm thick walls) was constructed to shield the detectors from the intense background radiation generated by the accelerator. A brass collimator with a 3 mm diameter aperture allowed for the electron beam to enter the bunker. A 6 mm diameter brass exit collimator was on the back wall of the bunker. This element, visible in figure 2, was present because the bunker was constructed for a previous and different experiment. A total of 96 BeO-OSL chips were placed on the front face of a Plexiglas support slab measuring $25 \times 25 \times 1.3$ cm$^3$. Each individual chip measured $4.7 \times 4.7 \times 0.5$ mm$^3$ with a nominal density of 2.85 g/cm$^3$. Before irradiation, the detectors were packed in groups of 4 and covered with a 10 µm of aluminum foil to shield them from visible light. Concurrently, 49 TL detectors, equally distributed among MCP-6, MCP-7, and CaSO$_4$:Dy, were placed in small holes in the support slab at a depth of 3 mm. The use of materials with different responses for each beam component is pivotal in mixed field dosimetry. Each TL detector measured 5 mm in diameter and 1 mm in thickness. The uncertainties on the positioning of the detectors are considered negligible compared to the dimensions of the chips.

Two separate irradiations were performed. During the first one, the front face of the slab was placed at a distance of 28 cm from the entry collimator. The primary electron beam impacted directly on the slab, without hitting any of the OSL nor TL detectors directly. For the second irradiation, instead, the front face of the slab was set at 28.7 cm from the collimator. In addition a polyethylene (PE) moderator, $36 \times 31 \times 8$ cm$^3$, was placed inside the bunker, 6 cm before the slab, shielding it completely from the primary beam.

GAFChromic$^{TM}$ HD-V2 were placed behind the slab and, only for the second irradiation, also behind the PE moderator. Finally, 5 small CR-39 detectors were placed on the front face of the slab, with a thickness of 0.9 mm and density 1.3 g/cm$^3$, and each covering a surface of ~2 cm$^2$. Figure 1 schematically shows the placement of each detector on the slab. While figure 2 is a 3D representation of the entire experimental setup.

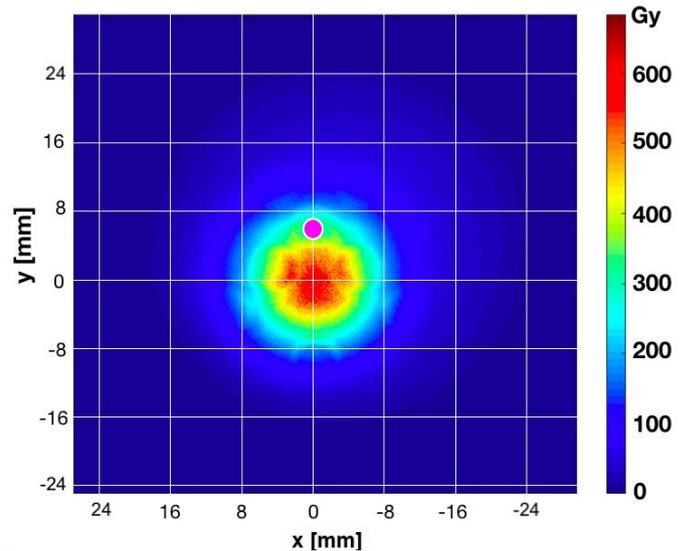

**Fig. 3** - Scan of the GAFChromic$^{TM}$ film placed behind the Plexiglas support slab during the first irradiation. Absorbed dose is displayed. The magenta dot marks the center of the slab shifted 0.75 cm above the beam axis. The circular areas of higher dose visible around the beam-spot are due to holes drilled in the slab to accommodate the TL detectors. These holes can be better appreciated in figure 5.

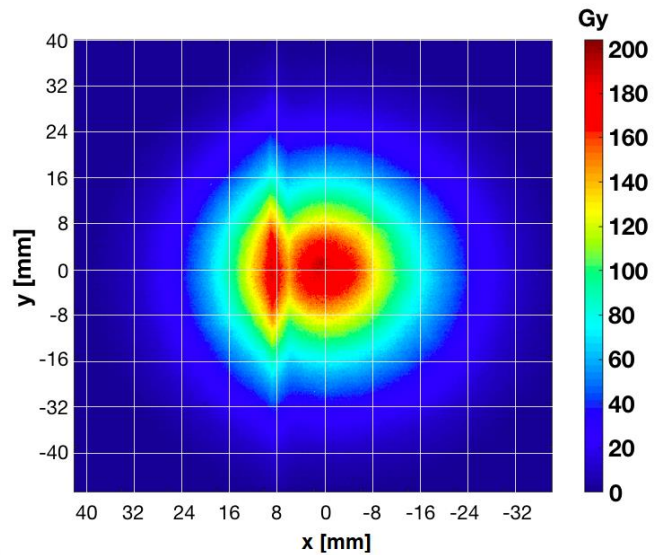

**Fig. 4** - Scan of the GAFChromic$^{TM}$ film placed behind the PE moderator during the second irradiation. Absorbed dose is displayed. The double-peak structure of the absorbed dose indicates a structural defect in the moderator.

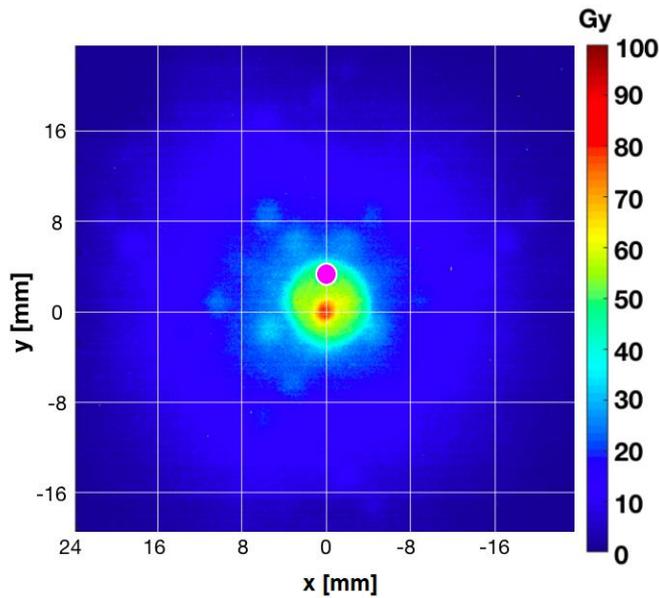

**Fig. 5 -** Scan of the GAFChromic™ film placed behind the support slab during the second irradiation. Absorbed dose is displayed. The magenta dot marks the center of the Plexiglas support slab shifted about 0.35 cm above the beam axis. The circular areas of higher dose clearly visible around the beam-spot are due to holes drilled in the slab supporting the dosimeters to accommodate TL detectors. This film, placed behind the Plexiglas slab, is sufficiently far away from the moderator (20cm) that the effect of the structural defect of the moderator on the radiation field is no longer visible.

## 4. Results and Discussion

Figures 3, 4, and 5 show the dose profile as measured with the GAFChromic™ films. From figure 3 and figure 5, we learn that the Plexiglas slab supporting the dosimeters was not centered on the beam axis. The center of the slab was estimated to be ~0.75 cm below the beam axis during the first irradiation and ~0.35 cm below during the second. Also, from high resolution scans it is clear that the beam spot was not circular, but elongated towards the upper right corner. This could be due to a misalignment of the slab and/or an intrinsic asymmetry of the beam itself. Figure 4, instead, clearly highlights a structural defect in the PE moderator: a narrow air gap between PE blocks. Furthermore, a visual analysis of high-resolution scans on sections of the film peripheral with respect to the beam show localized areas of higher dose. Thus weakness in the shielding-bunker made it non-radiation tight. The maximum absorbed dose measured by the GAFChromic™ on the back face of the slab, was 650 Gy and 75 Gy respectively for the two irradiations. The stray radiation field within the bunker was studied using BeO-OSL and various TL detectors. Data from these detectors were analyzed and compared amongst each other. Data are presented in figures 6 and 7. The presence of the moderator in the second irradiation clearly influences the dose distribution as a function of the distance from the beam.

The total delivered dose at the beam spot for irradiation 1 was much higher than in the second irradiation, but the beam was more collimated and less scattered because the moderator was not present. In fact, during the second irradiation, the moderator scattered the beam resulting in higher readings on the dosimeters placed about 4 cm or more from the beam axis.

It is important to notice that the BeO-OSL used are calibrated for air kerma, while the TL detectors are calibrated for absorbed dose. These two quantities are strongly related, but most definitely not the same and a direct numerical comparison of the results is not advisable. However, a qualitative comparison of data trends is possible. The 3 TL dosimeter types, instead, can be compared directly. The comparison was done among TL dosimeters placed in the same location (see figure 2). In both irradiations, among the TL dosimeters, the MCP-6 type shows, on average, the highest response. On average, they the measured values are 10% higher that the MCP-7 types at the same location. While, the $CaSO_4$:Dy dosimeters show, on average, the lowest response. The values shown in figure 7 are the total absorbed dose measured by each dosimeter. For a given irradiation, the relative decrease in dose as a function of the distance is comparable among the different detector materials. While, the difference in the absolute measured values can be justified by their different sensitivities to photons and thermal neutrons, by the uncertainty in the Plexiglas slab positioning which translates in an uncertainty on the position of the detectors, by the

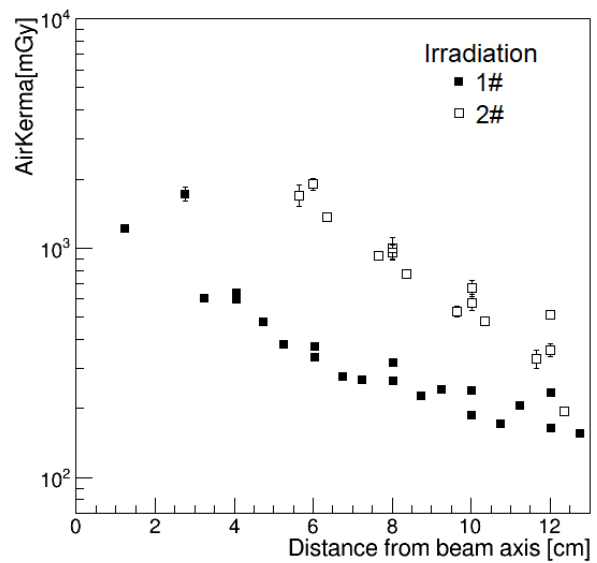

**Fig. 6 -** BeO-OSL detector responses to irradiation at the Microtron MT25. Error bars represent the statistical uncertainty on the measurement. Where not visible, error bars are small enough (better than 2%) to be completely contained in the markers. Values exceeding 2 Gy were omitted since the used detectors used were not calibrated beyond this values and therefore the obtained measurements would not be reliable.

asymmetry of the beam, and by the non-perfect radiation tightness of the bunker

Finally, the CR-39 detectors were read out. The results, in figures 8, show that the radiation field, measured by the CR-39 detectors, is uniform and not dependent on the distance from the beam axis. The measured radiation results from the interaction of the beam and the surrounding materials, mainly the front bunker wall. All 5 measurements are compatible within their uncertainties. The combined statistical and systematic uncertainties for each measurement is 20 %.

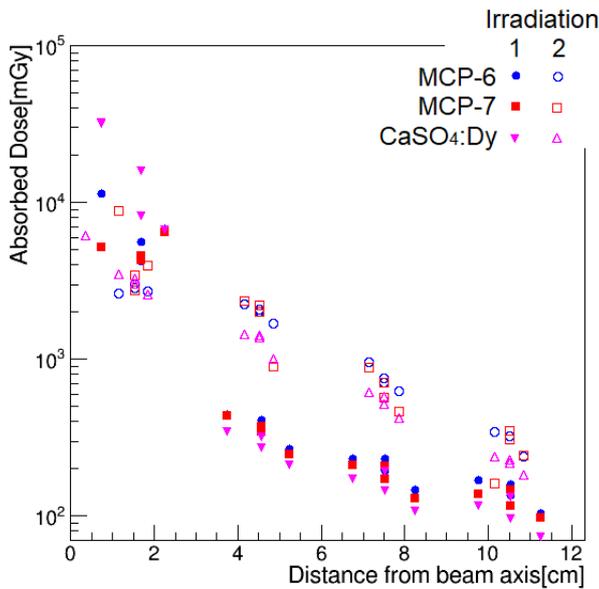

**Fig. 7** - TL detector responses to irradiation at the Microtron MT25. Error bars represent the statistical uncertainty on the measurement (5% relative uncertainty). Where not visible, error bars are small enough to be completely contained in the markers.

## 5. Conclusions

BeO-OSLs and MCP-6, MCP-7, and CaSO4:Dy TL dosimeters were studied in pulsed mixed radiation fields at the MT25 cyclic electron accelerator. This was a necessary and propaedeutic step for the subsequent development and validation of the dosimetry methodologies in laser driven and FLASH facilities. An initial inter-comparison among optical- and thermo- luminescent dosimeters was performed in pulsed mixed radiation fields at the Microtron MT25 in Prague, Czech Republic. The recorded discrepancies among the different detector types may be explained by the different detector sensitivities, non-uniformities in the geometrical setup, and structural defects of the shielding. The beam profile was analyzed using GAFChromic™ films, while the background due to neutrons and heavy charged particles was analyzed using CR-39 nuclear track detectors.

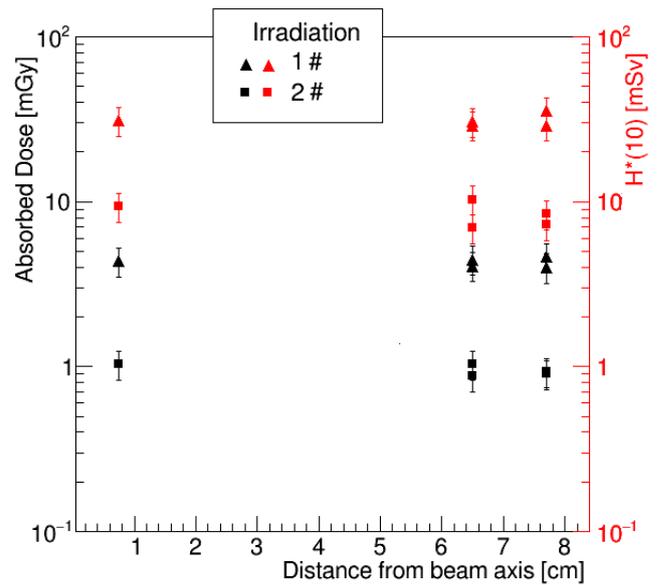

**Fig. 8** - Absorbed dose (black) and dose equivalent (red) measured by CR-39 nuclear track detectors. The triangular markers are the measurements from irradiation 1. While the squared one are for irradiation 2.

### Acknowledgements

This project 18HLT04 UHDpulse has received funding from the EMPIR programme co-financed by the Participating States and from the European Union's Horizon 2020 research and innovation programme.

### REFERENCES

[1] Binwei L, Feng G, Yiwei Y, Dai W, Yu Z, Gang F, Tangzhi D, Xiaobo D. FLASH Radiotherapy: History and Future. Front. in Oncol. 2021;11:1890, doi: 10.3389/fonc.2021.644400

[2] Favaudon V, Caplier L, Monceau V, Pouzoulet F, Sayarath M, Fouillade C, Poupon MF, Brito I, Hupé P, Bourhis J, Hall J, Fontaine JJ, Vozenin MC. Ultrahigh dose-rate FLASH irradiation increases the differential response between normal and tumor tissue in mice. Sci Transl Med. 2014 Jul 16;6(245):245ra93. doi: 10.1126/scitranslmed.3008973. Erratum in: Sci Transl Med. 2019 Dec 18;11(523): PMID: 25031268.

[3] Vozenin MC, De Fornel P, Petersson K, Favaudon V, Jaccard M, Germond JF, Petit B, Burki M, Ferrand G, Patin D, Bouchaab H, Ozsahin M, Bochud F, Bailat C, Devauchelle P, Bourhis J. The Advantage of FLASH Radiotherapy Confirmed in Mini-pig and Cat-cancer Patients. Clin Cancer Res. 2019 Jan 1;25(1):35-42. doi: 10.1158/1078-0432.CCR-17-3375.

[4] Loo BW, Schuler E, Lartey FM, Rafat M, King GJ, Trovati S, Koong AC, Maxim PG. (P003) Delivery of ultra-rapid flash radiation therapy and demonstration of normal tissue sparing after abdominal irradiation of mice. Int J Radiat Oncol Biol Phys 2017;98(2, Suppl.):E16, ISSN 0360-3016, doi: 10.1016/j.ijrobp.2017.02.101.

[5] Bourhis J, Sozzi WJ, Jorge PG, Gaide O, Bailat C, Duclos F, Patin D, Ozsahin M, Bochud F, Germond JF, Moeckli R, Vozenin MC. Treatment of a first patient with FLASH-radiotherapy. Radiother Oncol. 2019 Oct;139:18-22. doi: 10.1016/j.radonc.2019.06.019.

[6] Jaccard M, Durán MT, Petersson K, Germond JF, Liger P, Vozenin MC, Bourhis J, Bochud F, Bailat C. High dose-per-pulse electron beam dosimetry: Commissioning of the Oriatron eRT6 prototype linear accelerator for preclinical use. Med Phys. 2018 Feb;45(2):863-874. doi: 10.1002/mp.12713.

[7] Schueler E, Trovati S, King G, Lartey F, Rafat M, Loo B, Maxim P. (2016). TU-H-CAMPUS-TeP2-02: FLASH Irradiation Improves the Therapeutic Index Following GI Tract Irradiation. Medical Physics. 2016;43:3783-3783. doi: 10.1118/1.4957690.

[8] Montay-Gruel P, Acharya MM, Petersson K, Alikhani L, Yakkala C, Allen BD, Ollivier J, Petit B, Jorge PG, Syage AR, Nguyen TA, Baddour AAD, Lu C, Singh P, Moeckli R, Bochud F, Germond JF, Froidevaux P, Bailat C, Bourhis J, Vozenin MC, Limoli CL. Long-term neurocognitive benefits of FLASH radiotherapy driven by reduced reactive oxygen species. Proc Natl Acad Sci U S A. 2019 May 28;116(22):10943-10951. doi: 10.1073/pnas.1901777116.

[9] Diffenderfer ES, Verginadis II, Kim MM, Shoniyozov K, Velalopoulou A, Goia D, Putt M, Hagan S, Avery S, Teo K, Zou W, Lin A, Swisher-McClure S, Koch C, Kennedy AR, Minn A, Maity A, Busch TM, Dong L, Koumenis C, Metz J, Cengel KA. Design, Implementation, and in Vivo Validation of a Novel Proton FLASH Radiation Therapy System. Int J Radiat Oncol Biol Phys. 2020 Feb 1;106(2):440-448. doi: 10.1016/j.ijrobp.2019.10.049.

[10] Schüler E, Eriksson K, Hynning E, Hancock SL, Hiniker SM, Bazalova-Carter M, Wong T, Le QT, Loo BW Jr, Maxim PG. Very high-energy electron (VHEE) beams in radiation therapy; Treatment plan comparison between VHEE, VMAT, and PPBS. Med Phys. 2017 Jun;44(6):2544-2555. doi: 10.1002/mp.12233.

[11] Aßmann, R W, Grebenyuk J. Accelerator Physics Challenges towards a Plasma Accelerator with Usable Beam Quality, in Proc. 5th Int. Particle Accelerator Conf. (IPAC'14), Dresden, Germany, 2014 Jun. 961-964. doi: 10.18429/JACoW-IPAC2014-TUOBB01

[12] Ferrario M, Assmann R W. Advanced Accelerator Concepts. 2021. arXiv:physics.acc-ph, 2103.10843.

[13] Labate L, Palla D, Panetta D, Avella F, Baffigi F, Brandi F, Di Martino F, Fulgentini L, Giulietti A, Köster P, Terzani D, Tomassini P, Traino C, Gizzi L A. Toward an effective use of laser-driven very high energy electrons for radiotherapy: Feasibility assessment of multi-field and intensity modulation irradiation schemes. Sci Rep. 2020;10:17307. doi: 10.1038/s41598-020-74256-w

[14] Chiu C, Fomytskyi M, Grigsby F, Raischel F, Downer MC, Tajima T. Laser electron accelerators for radiation medicine: a feasibility study. Med Phys. 2004 Jul;31(7):2042-52. doi: 10.1118/1.1739301.

[15] Tajima T, Dawson T M. Laser Electron Accelerator. Phys Rev Lett, 1979;43(4):267-270 doi: 10.1103/PhysRevLett.43.267.

[16] Schüller A, Heinrich S, Fouillade C, Subiel A, De Marzi L, Romano F, Peier P, Trachsel M, Fleta C, Kranzer R, Caresana M, Salvador S, Busold S, Schönfeld A, McEwen M, Gomez F, Solc J, Bailat C, Linhart V, Jakubek J, Pawelke J, Borghesi M, Kapsch RP, Knyziak A, Boso A, Olsovcova V, Kottler C, Poppinga D, Ambrozova I, Schmitzer CS, Rossomme S, Vozenin MC. The European Joint Research Project UHDpulse - Metrology for advanced radiotherapy using particle beams with ultra-high pulse dose rates. Phys Med. 2020 Dec;80:134-150. doi: 10.1016/j.ejmp.2020.09.020.

[17] Petersson K, Jaccard M, Germond JF, Buchillier T, Bochud F, Bourhis J, Vozenin MC, Bailat C. High dose-per-pulse electron beam dosimetry - A model to correct for the ion recombination in the Advanced Markus ionization chamber. Med Phys. 2017 Mar;44(3):1157-1167. doi: 10.1002/mp.12111.

[18] Oancea C, Bălan C, Pivec J, Granja C, Jakubek J, Chvatil D, Olsansky V, Chiș V. Stray radiation produced in FLASH electron beams characterized by the MiniPIX Timepix3 Flex detector. JINST. 2022;17: C01003. doi: 10.1088/1748-0221/17/01/c01003

[19] Caresana M, Cassell C, Ferrarini M, Hohmann E, Manessi GP, Mayer S, Silari M, Varoli V. A new version of the LUPIN detector: improvements and latest experimental verification. Rev Sci Instrum. 2014 Jun;85(6):065102. doi: 10.1063/1.4879936.

[20] Zorloni G, Ambrožová I, Carbonez P, Caresana M, Ebert S, Olšovcová V, Pitzschke A, Ploc O, Pozzi, F, Silari M, Trompier F, Truneček R, Zelenka Z. Intercomparison of personal and ambient dosimeters in extremely high-dose-rate pulsed photon fields. Radiation Physics and Chemistry. 2020;172:108764. doi:10.1016/j.radphyschem.2020.108764.

[21] Karsch L, Beyreuther E, Burris-Mog T, Kraft S, Richter C, Zeil K, Pawelke J. Dose rate dependence for different dosimeters and detectors: TLD, OSL, EBT films, and diamond detectors. Med Phys. 2012 May;39(5):2447-55. doi: 10.1118/1.3700400. PMID: 22559615.

[22] Yukihara EG, McKeever SWS. Optically Stimulated Luminescence: Fundamentals and Applications. Wiley. 1st ed. 2011 ISBN: 978-0-470-69725-2P.

[23] Bulur E, Goksu H Y. OSL from BeO ceramics: new observations from an old material. Rad Meas. 1998;29(6):639-650.

[24] Cimmino A, Horváth D, Olšovcová V, Stránský V, Truneček R, Versaci R. Characterization of OSL dosimeters used at the ELI-beamlines laser-driven accelerator facility. J Radiol Prot. 2021 Dec 6;41(4). doi: 10.1088/1361-6498/ac14d5.

[25] Yukihara EG, Andrade AB, Eller S. BeO optically stimulated luminescence dosimetry using automated research readers. Radiat. Meas. 2016 94:27-34.

[26] Bos A J J. Theory of thermoluminescence. Rad Meas. 2006;41:S45-S56. doi: 10.1016/j.radmeas.2007.01.003.

[27] Bilski P, Olko P, Burgkhardt B, Piesch E, Waligorski M P R. Thermoluminescence efficiency of LiF:Mg,Cu,P (MCP-N) detectors to photons, beta electrons, alpha particles and thermal neutrons. Rad Prot Dos. 1994;55(1):31-38.

[28] Bilski P, Olko P, Budzanowski M, Ryba E, Waligorski M P R. 10 years of experience with high-sensitive LiF:Mg,Cu,P (MCP-N) thermoluminescent detectors in radiation dosimetry. Proceedings IRPA Regional Symposium, 1998;30(28):499-501.

[29] Moscovitch M, Personnel dosimetry using Li: Mg,Cu,P. Radiat. Prot. Dosim. 1999:85:49-56. doi: 10.1093/oxfordjournals.rpd.a032904.

[30] Olko P, Budzanowski M, Bilski P, Milosevic S, Obryk B, Ochab E, Simic M, Stegnar P, Waligorski M P R, S. Zunic Z. Application of MCP-N (LiF:Mg,Cu,P) TL detectors in monitoring environmental radiation. Nucl. Technol. Radiat. Prot., 2004:1:20-25. doi: 10.2298/NTRP0401020O.

[31] Yamashita T, Nada N, Onishi H, Kitamura S. Calcium sulfate activated by thulium or dysprosium for thermoluminescence dosimetry. Health Phys. 1971 Aug;21(2):295-300. doi: 10.1097/00004032-197108000-00016.

[32] Guelev, M.G., Mischev, I.T., Burgkhardt, B., Piesch, E. A two-element CaSO4:Dy dosemeter for environmental monitoring. Radiat. Prot. Dosim. 51(1), 35–40 (1994).

[33] Cartwright G, Shirk E K, Price P B. A nuclear-track-recording polymer of unique sensitivity and resolution Nucl. Instrum. Meth. 1978;153:457-460.

[34] Pachnerova Brabcova K, Ambrožová I, Koliskova Z, Malusek A. Uncertainties in linear energy transfer spectra measured with track-etched detectors in space. Nuc Instr and Meth in Physics Rese. 2013;713. doi: 10.1016/j.nima.2013.03.012.

[35] Lewis D, Micke A, Yu X, Chan MF. An efficient protocol for radiochromic film dosimetry combining calibration and measurement in a single scan. Med Phys. 2012 Oct;39(10):6339-50. doi: 10.1118/1.4754797.

[36] Vognar M, Simane C, Chvatil D. Twenty years of microtron laboratory activities at CTU in Prague. Acta Polytech. 2003:46;50-58. doi: 10.14311/424.

[37] Kapitza S P, Melekhin V N. The Microtron, Harwood Academic Publishers. London Chur; 1978.